# WHERE WAS MEAN SOLAR TIME FIRST ADOPTED?

**Simone Bianchi**
INAF-Osservatorio Astrofisico di Arcetri,
Largo E. Fermi, 5, 50125, Florence, Italy
simone.bianchi@inaf.it

**Abstract:** It is usually stated in the literature that Geneva was the first city to adopt mean solar time, in 1780, followed by London (or the whole of England) in 1792, Berlin in 1810 and Paris in 1816. In this short paper I will partially revise this statement, using primary references when available, and provide dates for a few other European cities. Although no exact date was found for the first public use of mean time, the primacy seems to belong to England, followed by Geneva in 1778–1779 (for horologists), Berlin in 1810, Geneva in 1821 (for public clocks), Vienna in 1823, Paris in 1826, Rome in 1847, Turin in 1849, and Milan, Bologna and Florence in 1860.

## 1 INTRODUCTION

The inclination of the Earth's axis with respect to the orbital plane and its non-uniform revolution around the Sun are reflected in the irregularity of the length of the day, when measured from two consecutive passages of the Sun on the meridian. Though known since ancient times, the uneven length of true solar days became of practical interest only after Christiaan Huygens (1629 –1695) invented the high-accuracy pendulum clock in the 1650s. For proper registration of regularly-paced clocks, it then became necessary to convert true solar time into mean solar time, obtained from the position of a fictitious mean Sun; mean solar days all having the same duration over the course of the year. Tables providing the equation of time, i.e. the difference between mean and true (sometimes called apparent) solar time, throughout the year, were computed already by Huygens, and in the early 1670s they were perfected by John Flamsteed (1646–1719, who was to become the first Astronomer Royal at the Royal Observatory, Greenwich- see, e.g., Turner, 2015) into the form used today.

At first, the use of mean solar time was of concern for astronomers and horologists only. Public clocks continued to be regulated on the true Sun, using sundials and meridian lines. Thus, it was necessary to register them often, at the risk of damaging the mechanisms. The increasing need to know the time precisely, together with the availability of higher-accuracy watches, eventually made the adoption of mean solar time necessary. But where was mean solar time first adopted?

In the modern literature it is often stated that Geneva was the first city to use mean solar time for public clocks, in 1780, followed by England (or just London, in some texts) in 1792, Berlin in 1810 and Paris in 1816: see, for example, Howse (1980: 82) and Dohrn-van Rossum (1996: 346). Their source is likely Bigourdan (1914: B.8). However, there is no reference to the source of that information in Bigourdan (1914), nor in earlier occurrences of the sequence, a few articles describing time zones around the years of their introduction (Möller, 1891; Oppermann, 1890; Rocca, 1893). References are instead provided in Houzeau (1882: 151), which is probably the main source for the later texts. The lack of primary sources in most texts was noted by Lundmark (1996), who questioned the date for the adoption of mean solar time in England and suggested, tentatively, that it happened later, together with the country-wide adoption of Greenwich mean time (I maintain here that it happened earlier than 1792). Even in the nineteenth century, Raab (1889) lamented how difficult it was to find information about this topic, because of the scant official and scientific documentation.

I recently had the opportunity to study the introduction of mean solar time (henceforth 'mean time') in Italy and compare it in a European context (Bianchi, 2019; 2020b). Unlike previous researchers, I could make use of the ever-increasing availability of on-line digital libraries. Thus, I was able to access some new primary sources, which has led me to revise some of the dates when mean time was introduced throughout Europe, and I present this new information.

## 2 FIRST GENEVA OR ENGLAND?

The quoted primacy in the use of mean time by Geneva fits with the importance of watch-making in the economy of the city. Indeed, it was the local *Société pour l'Encouragement des Arts et de l'Agriculture* that installed a mean time meridian on a wall of the Cathedral of St. Pierre; the aim was to help local watch-makers check the accuracy of their products, without the need to refer to tables for the equation of time. An analemma, an 8-shaped curve marking the Sun's position at noon of mean time throughout the year, was traced on top of an existing meridian by the astronomer Jacques-André Mallet (1740–1790), founder and Director of Geneva Observatory. The revised meridian was installed in 1778. The Geneva Government found the enterprise commendable, reimbursed the *Société* for the cost of the installation, and put a person in charge of ringing a bell at noon, so that artisans in the neighborhood could know mean time without the need of leaving their workshops. The bell signal was already at work in 1779, when it was put under the care of a church warden (Gautier, 1894; *Memoires*, 1780; Rambal, 1889). The service was intended for the watch-making industry, and not for public use at large (although the bell could be heard by everybody). Clocks in Geneva, including the Cathedral one, still remained regulated to true time. Mean time was eventually adopted a few decades later, starting from 15 April 1821. The day was chosen *ad hoc*, since it is one of the four in a year in which true and mean time coincided (i.e. the equation of time was null). The bell signal for the horologists remained in use for more than 20 years after then, until the late 1840s (Gautier, 1894).

The Director of Paris Observatory, Jérôme Lalande (1732–1807) quoted the *Memoires de la Société* (1780) in a note of his treatise *Astronomie* (Lalande, 1792: 341) and wrote that Geneva used mean time from 1780. He had added that information also in an earlier encyclopaedical entry about meridians (Lalande, 1785). In both references, however, he made clear that Geneva was not first: mean time was already in use in England before 1785. Referring to Lalande (1792), Houzeau (1882: 151) said that mean time "... was used in England in 1792." In later texts, this became "... it was used in England *from* 1792 ...", confusing the year in which Lalande's *Astronomie* was published with the beginning of the service.

The adoption of mean time in England must have occurred between the end of the seventeenth century and the mid-eighteenth century, but I could find no precise date in the literature. Later accounts said that public clocks in London "... never had shewn any other but mean time ..." (Vulliamy, 1828: 12). The fact that mean time was already in use for so long is also suggested by the first edition of *The Nautical Almanac* ... (Maskelyne, 1766). In fact, the ephemerides in the Almanac were provided in true time, in order to be readily compared with time measurements from the height of the Sun and used in the determination of geographical coordinates at sea. However, a caveat was added to the explanations:

> This [true] Time is different from that shewn by Clock and Watches well regulated at Land, which is called equated or mean Time ... that which should be shewn by a good Clock or Watch ... an equable Motion, such as that of Clocks and Watches ought to be. (The Nautical Almanac and Astronomical Ephemeris, 1766: 145, 150).

Indeed, Giuseppe Piazzi (1746–1826), Director of Palermo Observatory, stated that mean time had been used for public clocks in England prior to 1750 (Piazzi, 1798; see also Tuscano, 2016). Despite the lack of a precise year, it is nevertheless clear that the primacy for the public use of mean time belongs to England.

## 3 BERLIN AND VIENNA

In 1787 a clock made by the court and city horologist Christian Möllinger (1754 –1826) was installed over the entrance of the Berlin Academy of Sciences. The clock had two quadrants, one on the interior of the building, and one on the façade along the *Unter den Linden* boulevard and accessible to the public. Both quadrants had two hands for minutes, one showing true time and the other mean time. While the clock quickly became a popular stop-over for setting pocket watches, the presence of two hands for minutes confused the public and caused protests. After just a few weeks the mean time minute hand was removed from the external quadrant, which then only showed true time. At the same time, local authorities failed to get public clocks in accord with the Academy clock (Görike and Kiesant, 2021; Sauter, 2007).

Eventually, mean time was adopted for public use at the end of 1810, as usually reported in the literature. The switch was proposed by the Director of Berlin Observatory, Johann Elert Bode (1747–1826) at a meeting of the Academy of Sciences early in December that year. Soon after, the Academy clock was set on mean time, and the authorities had all public clocks regulated on that, and "... since then Berlin, as already other big cities did, relies on mean time ...." (Bode, 1811: 230). While it is not clear to which big cities Bode referred (unless he just meant London and Geneva), other cities in the Kingdom of Prussia must have followed the capital; from the edition for 1830, the *Berliner Kalender* (1829: xliii) started to use mean time for its ephemerides, "... since now in Berlin and in the most important Prussian cities the clocks are regulated on mean time ..."

Another astronomer, the Director of Vienna Observatory Joseph Johann von Littrow (1781–1840) was responsible for the switch in the capital of the Austrian Empire. From 1 March 1823, a bell at Vienna Observatory was first rung two minutes before noon, to alert clock regulators, and then every 2 seconds from 22 seconds before noon; at the last stroke, the bell of the Cathedral of St. Stephan rang, and on this all bell-towers and public clocks were to be regulated. Unlike previously, noon was that of mean time (Littrow, 1823).

## 4 PARIS

When speaking of the use of mean time in Geneva and England, Lalande (1785: 385) added: "... the Englishmen are surprised [to know] that true time, or solar time, is still used in France, despite its irregularities." The Court and Navy Horologist Ferdinand Berthoud (1727–1807) had already proposed the use of mean time in France, from as early as 1754. During the Revolution, Berthoud (1797: 3) made his proposal again: he hoped that the Académie des Sciences, which had made "... weights and measures uniform and invariable ...", would have encouraged the use of the uniform, mean time in place of the uneven true time. While noting that some nations already used mean time (though he did not say explicitly England and Geneva), Berthoud appealed to French pride and concluded

> ... that in a nation well known for Enlightenment, only mean time can be adopted for civil use ... (Berthoud, 1797: 11).

His proposal was backed by Lalande, who, in the *Connaissance des Tems* (1797: 187), stated "... that one should leave true time and use, even in society, mean time." A few years later, Berthoud (1802) reiterated his proposal, but to no avail. After more than a decade, a text printed in Dublin said "... we [in Ireland] reckon by mean time; in France they use solar time." (Ennis, 1816: xvii).

The statement that mean time started to be used in Paris in 1816 comes from the treatise *Astronomie Populaire* (Arago, 1854; see also Houzeau, 1882), a posthumous publication of the lectures of the Director of Paris Observatory, François Arago (1786–1853). Besides the year, the text does not provide many details on the implementation of the measure. It says only that the *Bureau des Longitudes* was asked by the prefect of the Seine Department, Gaspard de Chabrol (1773– 1843), for a report on the possible effects of the change: apparently, the prefect feared there would be social unrest; instead, the change passed unnoticed. The quoted year, however, must have been a mis-print. The change actually happened on 24 December 1826, when all Paris clocks, starting with the new clock on the building of the Paris *Bourse* (stock exchange) and that of the *Hôtel de Ville* (city hall), switched to mean time. As with Geneva in 1821, the choice fell on a date in which the equation of time was null. An article announcing the measure in the newspaper *Le Courier Français* commented that the administration took a laudable decision, finally putting Paris on a par with other cities, such as London, Amsterdam and Geneva (Du temps vrai, 1826). That the switch took place in 1826 is also confirmed by Vulliamy (1828). The same article includes a few sentences from the report that Arago made on behalf of the *Bureau des Longitudes*, stating again the utility of public mean time for the watch-making industry (similar words also are found in Arago, 1854).

## 5 ITALY

The switch to mean time happened later in Italy. In Rome, it was made shortly after the change in reckoning the hours. Until the middle of the eighteenth century, in the various Italian states the day (of 24 equal hours) started half an hour after sunset. This implied that clocks had to be regulated every day, but for practical reasons this was done every fortnight, using precomputed tables (Colzi, 1995).

From 1750, the Grand Duchy of Tuscany was the first state to adopt the hour count following the so-called French (or ultramontane) style (Bianchi, 2019), i.e. the usual system with a 24-hour day starting at midnight and having noon at 12 o'clock (of true time). The adoption in Italy was gradual, and was only complete by the end of the Napoleonic wars. After the Restoration, only the Papal States switched back to the old system but returned to the modern one after the ascent to the papacy of Pio IX in 1846. On the advice of Francesco de Vico S.J. (1805–1848), Director of Collegio Romano Observatory, the change was made more drastic, with the introduction of mean time and the start of a time signal on 1 January 1847 (Colzi, 1995; Secchi, 1877).

Another astronomer, Giovanni Antonio Amedeo Plana (1781–1864) was behind the switch to mean time of Turin, capital of the Kingdom of Sardinia (he was the Director of the local observatory); this happened in autumn 1849 (Il tempo vero e il tempo medio, 1849). Mean time was used afterwards for telegraphy and railways in the Kingdom of Sardinia, and later this practice extended to the Italian

Peninsula while the unification into the new Kingdom of Italy proceeded. The provisional governments of Milan, Bologna and Florence, for example, introduced the use of mean time for public clocks in 1860 (on 29 February, 20 March, and 24 December, respectively; Bianchi, 2019).

In Florence the use of mean time was hoped for in order to follow what was done "... in Turin and in every learned city of Europe." (Ridolfi, 1860). Before the adoption of mean time, the civic clock on the tower of the *Palazzo Vecchio* in Florence was set on true time by using a meridian line in the square below. Instead, from 24 December 1860 the noon of mean time was signaled to the *Palazzo Vecchio* by the lowering of a flag at the old Florence Observatory (the *Specola*; Bianchi, 2020a; 2020b). The public, believing that the Sun's indication gave the correct time, started to complain that the tower clock was no longer accurate (Bianchi, 2019). The authorities solved the problem by installing a new meridian line with an analemma for mean time (and this still exists today). It was traced by the observatory's Director, Giovanni Battista Donati (1826–1873).

## 6 CONCLUDING REMARKS

I have revised here the chronology of the adoption of mean time in Europe. Mean time was first used for public clocks in England, from at least the mid-eighteenth century; in Geneva from 1778–1779 (but only for the use of watch-makers, while in 1821 for public clocks); in Berlin in 1810; in Vienna in 1823; in Paris in 1826; in Rome in 1847; in Turin in 1849; in Milan, Bologna and Florence in 1860.

Since the scope of my work was mainly that of checking a statement common in the literature (like, e.g., in Howse, 1980), my research is in no way complete. Indeed, I found indications for the switch to mean time in other cities and countries. For these, however, I have not been able to find detailed references. I list them here, hoping they could be useful for further investigations: the whole of Ireland passed to mean time before 1816, and Amsterdam before 1826 (Section 4); most cities in Prussia between 1810 and 1829 (Section 3); Sweden, according to almanacs, in the early-1840s (Lundmark, 1996; Möller, 1891); Naples sometime before Rome, apparently the first city to do so in Italy (Decuppis, 1853); Bern and the whole of Switzerland in 1853 (Wolf, 1872).

## 7 NOTES

I am responsible for all French-to-English, German-to-English and Italian-to-English translations in this paper.

## 8 ACKNOWLEDGEMENTS